\documentstyle[epsf,10pt]{article}

\advance\textheight by \topskip

\begin{document}
\newcommand{\myover}[2]{ \left( \!\! \begin{array}{c} #1 \\ #2
                                     \end{array} \!\! \right) }
\newcommand{\smover}[2]{
   ( \!\! \begin{array}{c} \mbox{\small $#1$} \\[-2mm]
                           \mbox{\small $#2$} \end{array} \!\! ) }

\begin{titlepage}
\title{Site percolation and random walks
       on $d$-dimensional Kagom\'e
       lattices}

\author{Steven C. van der Marck \\
         SIEP Research and Technical Services \\
         P.O.~Box~60, 2280~AB~~Rijswijk, The Netherlands}

\end{titlepage}
\maketitle

\begin{abstract}
    The site percolation problem is studied on 
    $d$-dimensional generalisations of the Kagom\'e lattice.
    These lattices are isotropic and have the same coordination
    number~$q$ as the hyper-cubic lattices in $d$ dimensions,
    namely $q=2d$.
    The site percolation thresholds are calculated numerically
    for $d=$~3, 4, 5, and~6.
    The scaling of these thresholds as a function of
    dimension~$d$, or alternatively~$q$,
    is different than for hypercubic lattices:
    $p_c \sim 2/q$ instead of $p_c \sim 1/(q-1)$.
    The latter is the Bethe approximation, which is usually
    assumed to hold for all lattices in high dimensions.
    A series expansion is calculated, in order to understand
    the different behaviour of the Kagom\'e lattice.
    The return probability of a random walker on these
    lattices is also shown to scale as~$2/q$.
    For bond percolation on $d$-dimensional diamond lattices
    these results imply $p_c \sim 1/(q-1)$.
\end{abstract}


\section{Introduction}
The Kagom\'e lattice is one of the most interesting lattices
in two dimensions. It is one of the eleven Archimedean tiling
lattices, where all the vertices are of the same type
(see e.g.\ Weisstein 1997).
In the case of the Kagom\'e lattice each vertex touches
a triangle, hexagon, triangle, and a hexagon.
All these polygons are regular.
Moreover, the Kagom\'e lattice is closely related to
the other lattices in two dimensions.
The sites of the Kagom\'e lattice correspond to the
bonds of the honeycomb lattice, which in turn is the
dual of the triangular lattice.
Therefore, since the bond percolation threshold of
the honeycomb lattice is $1-2\sin(\pi/18)$,
the site percolation threshold of the Kagom\'e
lattice is $1-2\sin(\pi/18)=0.6527036\ldots$ too
(Sykes and Essam 1964).
The bond percolation threshold is not known exactly,
but has been calculated numerically with
high precision to be $ 0.524 \,405 \,3 \pm 0.000 \,000 \,5$
(Ziff and Suding 1997).

Although these percolation thresholds have been known
for quite some time, it is not clear why this site
percolation threshold is high, compared to other
lattices.
For instance, it is much higher than the threshold
$ 0.592 \,746 \,0 \pm 0.000 \,000\,5 $ for
the square lattice (Ziff and Sapoval 1986, Ziff 1992),
although its coordination number
$q=4$ is equal to that of the Kagom\'e lattice.
What is more striking is that even the pentagonal
lattice, which has a low average coordination number
of $q=3\frac{1}{3}$, has a lower site percolation
threshold $ 0.6471 \pm 0.0006 $ than the Kagom\'e
lattice (Van der Marck 1997).
In other words, the site percolation thresholds are
not ordered according to the coordination number~$q$.
This runs contrary to common intuition, which
leads one to expect that a lattice with a higher
connectivity has a lower percolation threshold.
If one searches for general formulas that correlate
percolation thresholds with dimension and
coordination number, the Kagom\'e lattice therefore
poses a problem.
Galam and Mauger (1996) introduced different classes
of lattices to avoid this problem,
and they used different correlations for these classes.
Although this enabled them to derive good correlations,
it prompts the question why certain lattices
belong to one class and others to another class.

An analogue in three dimensions was found recently:
there is a lattice with $q=6$ and percolation threshold
$ 0.3898 \pm 0.0008 $ (Van der Marck 1997b).
Compared to the simple cubic lattice, which also has $q=6$,
but a threshold of $ 0.3114 \pm 0.0004 $,
this threshold is much higher.
It is even higher than the threshold for several
lattices with coordination number~$q=5$ (Van der Marck 1997).

In this paper, a generalisation of the Kagom\'e
lattice to higher dimensions is given (Sec.~\ref{sec_kagome}),
and numerical calculations of the site percolation
thresholds for 3, 4, 5, and 6 dimensions are presented
(Sec.~\ref{sec_pc}).
The scaling of these thresholds as a function of dimension
appears to be different than for hypercubic lattices.
For the latter, Gaunt, Sykes and Ruskin (1976)
calculated a series expansion in $1/(2d-1)$,
where $d$ is the number of dimensions.
The leading term in their series is $p_c(d) = 1/(2d-1)$,
which is the so-called Bethe approximation.
This approximation holds exactly for Bathe lattices
(see e.g.\ Stauffer and Aharony 1992).
For hypercubic lattices, the approximation underestimates
the percolation threshold in low dimensions, but
improves in accuracy in higher dimensions.
In Sec.~\ref{sec_exp} the series expansion for
the $d$-dimensional Kagom\'e lattices is studied.
It is suggested that in this case the leading
term in the series is $1/d$, not~$1/(2d-1)$.

The return probability of a random walker on
$d$-dimensional Kagom\'e lattices is discussed
in Sec.~\ref{sec_walk}.
Ishioka and Koiwa (1978) conjectured
that this probability is a good
estimator for the percolation threshold.
Indeed it is shown here that this return probability
also scales as~$1/d$.
Section~\ref{sec_discus} contains a discussion of the results.
These results give some insight into the
problem why the 2-dimensional Kagom\'e lattice
has a high site percolation threshold.

\section{Kagom\'e lattices in $d$~dimensions}
\label{sec_kagome}
The Kagom\'e lattice can be defined in $d$~dimensions
as follows.
The lattice has a $(d+1)$-point basis, and these

points form a regular $d$-dimensional polytope.
All the points of this basis are direct neighbours
of each other.
Let us denote the basis points by~${\bf b}_i$,
for~$i=0,\ldots,d$.
The lattice can be constructed by translation of
the basis in $\frac{1}{2}d(d+1)$ directions.
These translations are given by the vectors
$2({\bf b}_j - {\bf b}_i)$ for $j \neq i$.
This is a dependent set of vectors.
One can select a minimal set of $d$~vectors
by setting e.g.~$i=0$,~$j=1,\ldots,d$.

A site of the lattice can be identified by
its number~$i$ within the base polytope
(runs from $0$ to $d$), and the translation~${\bf x}$
with respect to a reference position.
Consider a site $\{ i, {\bf x} \}$.
This site has $d$~neighbours in the same
polytope, $\{ j \neq i, {\bf x} \}$, and
another $d$~neighbours in adjacent polytopes,
$\{ j \neq i, {\bf x} + 2( {\bf b}_i - {\bf b}_j ) \}$.
Therefore the coordination number of this lattice
is~$q=2d$.
Also, because there is no preferential direction
in the construction,
all the directions are equivalent for this lattice,
i.e.\ it is an isotropic lattice.

So the Kagom\'e lattice
resembles the cubic lattice in the sense
that both are isotropic $d$-dimensional lattices
with coordination number~$q=2d$.
However, we already know that in two and three dimensions,
the site percolation thresholds of these lattices
are distinctly different.

In two dimensions the site
percolation problem on the Kagom\'e lattice
is equivalent to the bond percolation problem
on the honeycomb lattice (Sykes and Essam 1964).
This was shown by means of the
star-triangle transformation.
Analogous to the star-triangle transformation,
one can use a `star-tetrahedron' transformation
in three dimensions, see Fig.~\ref{fig_startria}.
The site percolation problem on the tetrahedron~ABCD,
i.e.\ the three-dimensional Kagom\'e lattice,
is equivalent to the bond percolation problem
on the dashed lattice.
This dashed lattice is the diamond lattice,
for which $p_{c,b}=0.3893 \pm 0.0003$ has been
calculated (Van der Marck 1997).
Generalising to $d$~dimensions,
the site percolation on the
$d$-dimensional Kagom\'e lattice 
is equivalent to bond percolation on
the $d$-dimensional diamond lattice,
which has $q=d+1$.

The Kagom\'e lattice defines a
tiling of $d$-dimensional space.
In two dimensions, the Kagom\'e lattice defines
a tiling of the plane in terms of a regular triangle
and a regular hexagon.
One can construct this hexagon from the base triangle:
take three neighbouring base triangles,
as in Fig~\ref{fig_kagome}b.
These three form a larger regular triangle.
The hexagon appears when one truncates the larger triangle
by taking away the three smaller triangles.
In two dimensions this is a rather complicated
description of the tiling, but the advantage is that
one can use an identical procedure in $d$~dimensions.
One can start with the regular base polytope of $(d+1)$~points
in $d$~dimensions, ${\bf b}_i$ (for~$i=0,\ldots,d$).
This polytope is shifted in $d$~directions by
the vectors $2( {\bf b}_i - {\bf b}_0 )$ for $i=1,\ldots,d$.
This defines a larger regular polytope, which
is then truncated by taking away the $(d+1)$ small
polytopes.
In three dimensions, e.g., the resulting polyhedron
is the truncated tetrahedron~(see e.g.\ Weisstein 1997).
The $d$-dimensional space is filled with the base regular
polytope and the larger truncated polytope.

In view of the considerations in sections \ref{sec_exp}
and~\ref{sec_walk}, it is helpful to discuss
one more important property
of the $d$-dimensional Kagom\'e lattice:
two adjacent sites have $d-1$ {\it common neighbours}.
This is demonstrated by the following arguments.
When the two sites are within the same base
polytope, characterised by
$\{ i, {\bf x} \}$ and $\{ j, {\bf x} \}$ ($j\neq i$),
all the other $d-1$~sites of that polytope are
common neighbours.
When the two sites are {\it not} within the
same base polytope, they are characterised by
$\{ i, {\bf x} \}$ and
$\{ j, {\bf x} + 2( {\bf b}_i - {\bf b}_j) \}$.
Note that $j$ cannot be equal to~$i$ for two adjacent sites.
The $d-1$ common neighbours of these two sites are
$\{ k, {\bf x} + 2( {\bf b}_i - {\bf b}_k) \}$
with the restriction that~$k \neq i, j$.
This is because they are one step ${\bf b}_i-{\bf b}_k$
away from $\{ i, {\bf x} \}$, and
one step ${\bf b}_k-{\bf b}_j$ from
$\{ j, {\bf x} + 2( {\bf b}_i - {\bf b}_j) \}$.

\section{Percolation thresholds}
\label{sec_pc}
The site percolation thresholds of
the Kagom\'e lattice in 3, 4, 5, and 6~dimensions,
are show in Table~\ref{tab_big}
as a function of the lattice size.
I used the method given by Stauffer and Aharony (1992).
This method is simple to program, but in its simplest form
it has the drawback that one needs to have all
$N=(d+1)L^d$ sites of a network resident in memory.
Therefore the calculation is restricted to relatively
low values of~$L$, the linear size of the lattice.
Especially in higher dimensions this is rather restrictive:
the highest $L$-value used was $L=14$ for the
$6$-dimensional Kagom\'e lattice.
One can check in a scaling plot, Fig.~\ref{fig_scaling},
that the used lattices are indeed large enough to be
in the scaling regime, and hence allow extrapolation to
infinite lattice sizes.

For each of the lattices, percolation in only one direction,
say the ${\bf x}_1$-direction, was checked.
The lattice was defined to be percolating whenever
the two boundaries
$\{ i=0, {\bf x}_1 = 1 \}$ and $\{ i=d, {\bf x}_1 = L \}$
were connected.
The boundary conditions in the other directions
were not periodic.

The percolation thresholds of a lattice of size~$N$
obey the scaling relation
\begin{equation}
  \left| p_c(N) - p_c(\infty) \right|  \sim  N^{-1/(\nu d)} .
  \label{eq_scaling}
\end{equation}
Here the critical exponent~$\nu$ is $0.88$ in three dimensions and
$0.68$, $0.57$, $0.5$ in 4, 5, and 6 dimensions respectively
(Stauffer and Aharony 1992).
The results quoted in the row marked~$\infty$ in
table~\ref{tab_big} are fits of~$p_c(N)$ to this scaling relation,
using the last three points in the table.

The data and the fits are shown in Fig.~\ref{fig_scaling}
for $d=3$ and $d=6$.
From the figure it can be concluded that the input values
for the fits are indeed within the scaling regime,
and that the fits are therefore sensible.
For $d=6$ the difference $p_c(N)-p_c(\infty)$ is larger than
for lower dimensions, because the number of lattice sites
that are near to a boundary increases with dimension.

Also shown in Table~\ref{tab_big} are results for the percolation
thresholds of the cubic lattice. These numbers agree with the
literature (Stauffer and Aharony 1992) and can be seen as a
check that the programs used in the present work are correct.

The difference between the percolation threshold for the Kagom\'e
lattices and the (hyper-) cubic ones increases with dimension.
This is illustrated in Fig.~\ref{fig_dimen}.
In the next section a series expansion is studied, in order
to understand this behaviour.

\section{Series expansion}
\label{sec_exp}
One could conjecture, based on the numerical estimates of the
percolation thresholds, that the scaling of these thresholds
as a function of dimension is different for the Kagom\'e lattice
than for the cubic lattice.
The scaling behaviour of the cubic lattice was studied
by Gaunt {\it et al.} (1976). Their result for the percolation
threshold was
\begin{equation}
   p_c(d) = \frac{1}{2d-1} \left( 1 + \frac{3/2}{2d-1} +
                  \frac{15/4}{(2d-1)^2} + \frac{83/4}{(2d-1)^3}
                  + \ldots \right) .
   \label{eq_GSR}
\end{equation}
The leading term is the percolation threshold
of a Bethe lattice with coordination number~$q=2d$.
Equation~\ref{eq_GSR} agrees well with the known numerical
estimates for the cubic lattices for $d \geq 3$, as is clear
from Fig.~\ref{fig_dimen}.

Gaunt {\it et al.} (1976) used cluster counting to calculate
a series expansion for the mean size~$S$ of clusters
at low probabilities~$p$.
Since the mean cluster size diverges at the percolation
threshold, they studied the radius of convergence
of the series expansion of~$S$.
They argued that this radius of convergence will be
determined predominantly by the singularity at~$p=p_c$,
which enabled them to derive Eq.~(\ref{eq_GSR}).
In this section these techniques
are used to study the percolation thresholds of
the Kagom\'e lattice in higher dimensions.
At the basis of the approach are the `perimeter polynomials',
as described by Sykes and Glen (1976).
Denoting the mean number per lattice site
of clusters of $s$~sites by $\bar{n}_s$, we have e.g.
\begin{eqnarray}
  \bar{n}_1  & = & p \; q^{2d} \;, \nonumber \\
  \bar{n}_2  & = & d \; p^2 \; q^{3d-1} \;.
  \label{eq_n12}
\end{eqnarray}
Here $p$ is the probability that a site is occupied, and
$q=1-p$ the probability that a site is empty.
One can interpret $p(s)=s\bar{n}_s$ as the probability that a site
is occupied by a cluster of size~$s$.
When one sums $p(s)$ over all cluster sizes,
one gets the probability that a site is occupied
by {\it any} cluster, i.e.~$p$:
\begin{equation}
  \sum_{s=1}^{\infty} s \; \bar{n}_s =
  \sum_{s=1}^{\infty} p(s)           = p.  \label{eq_identity}
\end{equation}
The power $3d-1$ of~$q$ in the expression for $\bar{n}_2$ indicates
that all clusters of 2~sites are surrounded by $3d-1$~neighbours.
Already we see a difference with the cubic lattice,
which has $\bar{n}_2 = d p^2 q^{4d-2}$.
So even though a site on the Kagom\'e lattice has as many
neighbours as a site on the cubic lattice (namely $2d$),
a $2$-site cluster has less neighbours on a Kagom\'e lattice
than on a cubic lattice.
This is because on the Kagom\'e lattice two adjacent sites
have $d-1$ common neighbours, whereas on the cubic lattice
they have none.

A relatively simple counting procedure by hand reveals
the next few perimeter polynomials to be
\begin{eqnarray}
  \bar{n}_3  & = & {\textstyle \frac{1}{3}} d (4d-1)
                   \; p^3 \; q^{4d-2} ,
                 \label{eq_n345} \\
  \bar{n}_4  & = & {\textstyle \frac{1}{12}} d (5d-1)(5d-2)
                   \; p^4 \; q^{5d-3} , \nonumber \\
  \bar{n}_5  & = & \left[ {\textstyle \frac{1}{60}}
                       d (6d-1)(6d-2) (6d-3) - 2d(d-1)
                   \right] p^5 \; q^{6d-4}
                 \nonumber \\
           &   & + 2d(d-1) \; p^5 \; q^{6d-5} . \nonumber
\end{eqnarray}
In Figs.~\ref{fig_count3}, \ref{fig_count4} and~\ref{fig_count5}
the contributing graphs are depicted in two dimensions.
In the expressions for $\bar{n}_1, \ldots, \bar{n}_4$ a clear
pattern seems to emerge.
The coefficient of $\bar{n}_s$ has factors $(s+1)d-1$, $(s+1)d-2$,
$\ldots$, and the power of~$q$ is always~$(s+1)d-(s-1)$.

However, this trend is broken in $\bar{n}_5$.
Although the sum of the two coefficients appearing here
still has the factors $(s+1)d-1$, etc., there are now terms
with different powers of~$q$.
In other words, in $\bar{n}_5$ we encounter for the first time
that there are clusters with different numbers of neighbours;
clusters with $6d-4$ neighbours and clusters with $6d-5$ neighbours.
The latter ones are depicted in Fig.~\ref{fig_snake}.
They are the type of cluster that `bites itself in the tail'.
At first sight, one would think that this would occur 
already for clusters of two sites, but in those cases
the number of neighbours is not affected.
This number is only affected for the clusters
depicted in Fig.~\ref{fig_snake}.
On the cubic lattice this occurs for the first time
for clusters of $3$~sites.
Another difference is that for the cubic lattice the leading
power of~$q$ is $2sd - 2(s-1)$, compared to $(s+1)d-(s-1)$
for the Kagom\'e lattice.

The identity (\ref{eq_identity}) can be exploited to determine
one more mean cluster number, namely~$\bar{n}_6$,
albeit in the limit of~$q \rightarrow 1$.
Following Sykes and Glen (1976) I substitute
Eqs.~(\ref{eq_n12}) and~(\ref{eq_n345}) into the
identity~(\ref{eq_identity}), and set $q=1-p$.
This yields a power series in~$p$, of which the coefficient of
$p$ is~$1$, and the coefficients of $p^2, \ldots, p^5$ vanish
identically.
The higher order coefficients do not vanish, because the number
of clusters of six or more sites were not included.
Because $\bar{n}_6$ is the only missing term that can contribute terms
of order $p^6$, we can conclude
\begin{equation}
  \left. \bar{n}_6 \right|_{q \rightarrow 1} =
    \left[
     \frac{1}{360} d(7d-1)(7d-2)(7d-3)(7d-4) - \frac{5}{3} d(d-1)
    \right] p^6 .
\end{equation}
The mean size of clusters at low probabilities, $S$,
is defined as
\begin{equation}
  S =  \frac{\sum_s s \; p(s)}{\sum_s p(s)} =
       \frac{1}{p} \sum_s s^2 \; \bar{n}_s .
\end{equation}
Again Eqs.~(\ref{eq_n12}) and~(\ref{eq_n345}) can be used, and
$q=1-p$, to derive a power series
\begin{eqnarray}
  S(p) & = & 1 + 2(dp) + 2(dp)^2 + 2(dp)^3 + 2(dp)^4 \nonumber \\
       &   & \hphantom{1 }
               + 2(dp)^5 \left[ 1 + \frac{10}{d^3} \left( 1 -
                                    \frac{1}{d} \right) \right] 
               + {\cal O}\left( p^6 \right) .
\end{eqnarray}
The first few terms in this low density expansion
of $S(p)$ are remarkably simple.
It is almost a geometric series, until the fifth power.
The extra terms in the coefficient of $(dp)^5$
are due to the clusters that `bite in their own tail'.

Although the resemblance of $S(p)$ to a geometric series is not
exact, it does suggest that $1/d$ is the obvious candidate for
expansion parameter.
This in contrast to the cubic lattice, where it is $1/(2d-1)$,
see Eq.~(\ref{eq_GSR}).
In fact, the resemblance to a geometric series suggests that
a singularity of $S(p)$ should occur in the vicinity of~$dp=1$.
Therefore one can expect the percolation threshold to scale as
\begin{equation}
   p_c \sim \frac{1}{d}.
   \label{eq_scalepc}
\end{equation}
In Fig.~\ref{fig_dimen} the relation $p_c=1/d$ is shown
with a dashed line.
The percolation thresholds for $d=5$ and~$6$ are
already reasonably well approximated by this relation.
In an attempt to calculate the percolation threshold
for $d=8$, I computed
$0.1059 \pm 0.0005$ for $L=5$,
$0.1086 \pm 0.0006$ for $L=6$, and
$0.1117 \pm 0.0005$ for $L=7$.
Based on the last two points one can fit the percolation
threshold to be $0.120 \pm 0.003$, which is close to the
value~$0.125$ one would expect on the basis
of Eq.~(\ref{eq_scalepc}).
However, the point for $L=5$ does not lie on the same fit,
indicating that these lattice sizes are not yet large enough.
Therefore there is probably also a small systematic error
in the determination of the value~$0.120$.

When one wants to refine the scaling behaviour given
by Eq.~(\ref{eq_scalepc})
with terms of the order of~$1/d^2$ and further,
analogous to Eq.~(\ref{eq_GSR}),
more terms in the series expansion would be required,
plus a careful mathematical analysis of the radius
of convergence of the series.
This is beyond the scope of the present paper.

\section{Random walks}
\label{sec_walk}
It is interesting to study the return probability~$P_r$
of a random walker on a $d$-dimensional Kagom\'e lattice.
Ishioka and Koiwa (1978) suggested
that $P_r$ is an upper bound for the site percolation
threshold on any lattice:~$P_r \geq p_c$.
For Bethe lattices, the equal sign holds
(Hughes and Sahimi 1982).
The arguments given by Ishioka and Koiwa to support
their conjecture are not exact, but
the relation appears to work fairly well.
$P_r$~and $p_c$ lie closer together,
the more connected a lattice is.
Also Sahimi {\it et al.} (1983) studied the relation between
a random walker (not self-avoiding) and percolation,
albeit bond percolation.

The return probability of a random walker can be
calculated numerically with a simple computer program.
One can let $N_w$~walkers perform at maximum $N_s$~steps,
and count the number of walkers that have re-visited
the site they started from.
Alternatively, one can let $N_w$~walkers perform
steps until they are either back at the origin or
further away from the origin than a certain predefined distance.
I have used both methods to estimate~$P_r$ with an
estimated inaccuracy of about~$0.001$.
Note that there are two sources of inaccuracy.
The first one is a statistical uncertainty,
which scales as~$1/\sqrt{N_w}$.
The second one is a systematic error, because each
walker is stopped at a certain moment (after $N_s$ steps
or at a given distance from the origin).
For each of these walkers there is a finite probability
that they would have reached the origin, when given
enough time.
As a result, the numerical estimates have a systematic
error to the downside. This bias can be made smaller by
using a large number of steps, or a large cut-off distance.
In the calculation of the numbers quoted
in Table~\ref{tab_big}, I used $N_s=10^6$ and higher,
and~$N_w=10^5$.
The results for the cubic lattices are consistent with
Finch (1997) and Flajolet (1995).

The numerical values for~$P_r$ are close to the
percolation thresholds, especially for~$d \geq 4$.
It looks as if the return probability of a random walker
obeys the same scaling relation as the percolation threshold.
In the remainder of this section I therefore calculate
a crude approximation for the return probability
using simple arguments.
This approximation corroborates the scaling of the
return probability as~$P_r \sim 1/d$.

Consider a random walker on the $d$-dimensional Kagom\'e lattice.
Since all directions of the lattice are equivalent,
it does not make a
difference which step the walker makes first.
Assume, without loss of generality, that the walker remains
within the base polytope~$\{ i, {\bf 0} \}$.
Assume further, as a first approximation, that he
stays within this polytope for a number of steps,
and then returns to the origin.
The walker can take an arbitrary number
of steps within this base polytope, with a
probability~$(d-1)/(2d)$
(the walker can choose from $2d$~directions, $d$~of which are
outside the base polytope, and one of the~$2d$ is the origin).
After a number of steps the walker should step back to
the origin, which happens with a probability~$1/(2d)$.
This approximation yields
\begin{equation}
  P_{r,1} = \sum_{s=2}^{\infty} \frac{1}{2d}
                              \left( \frac{d-1}{2d} \right)^{s-2}
          = \frac{1}{1+d} .
\end{equation}
This approximation already reveals an interesting
point: in high dimensions, the return probability scales
as $P \sim 1/d$.
Since the probability~$1/(1+d)$ is the exact probability
of return via a few selected paths, we also know that
the exact probability on return via {\it any} path
will be higher than~$1/(1+d)$.
It is therefore impossible that it is as low
as~$1/(2d-1)$, as is the case for cubic lattices.

One can improve on the above approximation by allowing
the walker to step outside the base polytope occasionally.
Choose for instance $i$~sites~$x_i$ from which the walker
steps outside.
If the walker makes $s$~steps within the base polytope
before returning to the origin, there
are $\smover{s-1}{i}$ possible choices (because
the last of the $s$~steps is to the origin):
\[
  P_{r,2} = \sum_{s=2}^{\infty} \frac{1}{2d}
                  \left( \frac{d-1}{2d} \right)^{s-2} \times
            \sum_{i=0}^{s-1} \myover{s-1}{i}
                  \left( \frac{1}{4d} \right)^i .
\]
The factor $1/(4d)$ emerges because the walker has a probability
of~$1/2$ to step outside the base polytope,
and a probability~$1/(2d)$ to immediately step back to~$x_i$.
One can also allow the walker to make a number of additional
steps, as long as he is only one step away from $x_i$ and
outside the base polytope.
There are $(d-1)$ possibilities out of a
total of $2d$ to make such a step.
\begin{eqnarray}
  P_{r,2} & = & \sum_{s=2}^{\infty} \frac{1}{2d}
                  \left( \frac{d-1}{2d} \right)^{s-2} \times
            \sum_{i=0}^{s-1} \myover{s-1}{i} \left[ \frac{1}{4d} 
            \sum_{k=0}^{\infty} \left( \frac{d-1}{2d} \right)^k
                  \right]^i \nonumber \\
          & = & \frac{1}{d} \cdot
                \frac{ 1+3/(2d) }{ 1+3/(2d)+3/(2d^2) }.
\end{eqnarray}
This type of reasoning can be taken a step further, yielding
\begin{equation}
   P_{r,3} = \frac{1}{d} \cdot \frac{ 1 + 2/d             + 9/(4d^2) }
                                    { 1 + 2/d + 11/(4d^2) + 9/(4d^3) }.
   \label{eq_pr3}
\end{equation}
All the paths that are included in this way are paths where
the walker returns to the origin through the base polytope he
started in.
The first contributions from paths that return from the
opposite side are paths of six steps, see Fig.~\ref{fig_snake}.
As there are $2d(d-1)$ of such paths, the probability to return
from the opposite side in six steps is~$(d-1)/(2d)^5$, which
is of the order~$1/d^4$.
Thus the conclusion that the return probability scales as~$1/d$
remains unchanged.

As a numerical check on this scaling behaviour, I calculated
the return probability for the $8$-dimensional Kagom\'e lattice:
$P_r = 0.126 \pm 0.001$.
This compares well with the approximation~$0.124$
from Eq.~(\ref{eq_pr3}).

\section{Discussion}
\label{sec_discus}
The Kagom\'e lattice has rather different properties
in $d$~dimensions than the cubic lattice:
it does not behave as a Bethe lattice in high dimensions,
although one usually assumes that all lattices do.
The series expansion of Section~\ref{sec_exp} provides
some insight into why this is the case.
One clear difference with the cubic lattice is that the
leading power of~$q$ in the expression for the number
of clusters of $s$~sites is $(s+1)d-(s-1)$
for the Kagom\'e lattice, but
$2sd - 2(s-1)$ for the cubic lattice.
The coefficient of~$sd$ differs by a factor~2 here.
The significance of this power of~$q$ is that it is
the `coordination number' of a cluster of $s$~sites,
just as $q$ is the coordination number of a single site.
The underlying reason for the different powers of~$q$
is that two adjacent
sites on the Kagom\'e lattice always have $d-1$~common
neighbours, whereas two adjacent sites on the
cubic lattice have none.
In high dimensions this difference does not disappear, but
instead becomes more important.

The same difference also explains why the return probability
is different for these lattices.
Since adjacent sites have $d-1$ common neighbours,
there is, in high dimensions, an increasing number of ways
to return to the origin in three steps.
On a Bethe lattice, such paths do not exist: here a walker
can only return to the origin by retracing his steps.
Therefore the Kagom\'e lattice does not
behave as a Bethe lattice in high dimensions.

Going back to the $2$-dimensional Kagom\'e lattice,
one could say that one of the reasons why it has a much
higher site percolation threshold than the square lattice,
is because of its `common neighbours'.

We can apply the same reasoning to {\it bond} percolation.
The bond percolation threshold of the Kagom\'e lattice is also
higher than that of the square lattice, but the difference is
smaller than for the site percolation threshold.
Two adjacent bonds on the square lattice always have $2$~common
neighbours.
On the Kagom\'e lattice, two adjacent bonds have, depending
on their relative position, $2$ or $3$~common neighbours.
The former happens in $\frac{2}{3}$ of the situations,
the latter in $\frac{1}{3}$.
Therefore, for {\it bond} percolation the Kagom\'e lattice
has more resemblance to the square lattice, but still the
two lattices are different.
This is consistent with the difference in bond percolation
thresholds being smaller, but not negligible.

Finally, let us consider the relation to
the diamond lattice in $d$~dimensions.
Since the site percolation threshold of the Kagom\'e lattice
scales as $p_c\sim 1/d=2/q$, this holds for the bond percolation
threshold of the diamond lattice as well.
But for the diamond lattice~$q=d+1$, so $p_{c,dia} \sim 1/(q-1)$,
which is the familiar Bethe result once more.
This highlights a more general property.
Each bond problem can be mapped onto a site problem
(see e.g.\ Kesten 1982).
When the bond problem is on a lattice with $q=q_b$,
the lattice for the corresponding site problem
will have $q_s=2(q_b-1)$.
As a consequence, for each class of lattices for which the
bond percolation thresholds scale as $p_c \sim f(q)$,
there is another class of lattices for which the
site percolation thresholds scale as
$p_c \sim f(\frac{1}{2}q+1)$.
It would therefore be more consistent to describe
bond percolation in terms of the number of neighbours
a bond has, which is~$2(q-1)$.

\section*{Acknowledgments}
I would like to thank
Ed Stephens for critically reading the manuscript and
Shell International Exploration \& Production for permission
to publish this paper.



\setlength{\unitlength}{1cm}

\begin{figure}[htb]
  \begin{center}
    \begin{picture}(7.0,6.0)(0,0)
      \epsffile{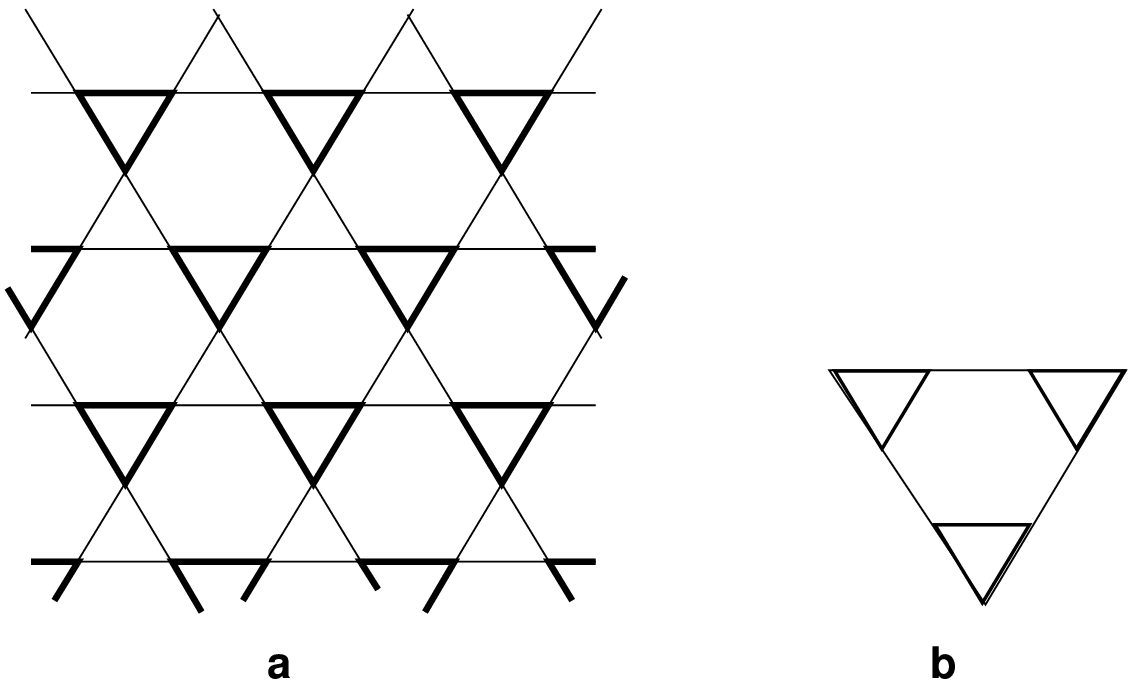}
    \end{picture}
  \end{center}
  \caption[]{(a) A description of the Kagom\'e lattice as
                 a lattice with a 3-point basis. The basis
                 points form a (regular) triangle.
             (b) Three base triangles define a larger triangle,
                 which truncates to a hexagon.
            }
  \label{fig_kagome}
\end{figure}

\begin{figure}[htb]
  \begin{center}
    \begin{picture}(7.0,6.0)(0,0)
      \epsffile{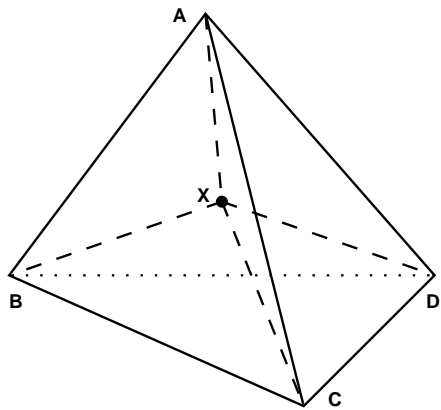}
    \end{picture}
  \end{center}
  \caption[]{The star-tetrahedron transformation.
             The dashed lines AX,$\ldots$,DX, that come
             together in the point~X, form the diamond lattice.}
  \label{fig_startria}
\end{figure}

\begin{figure}[htb]
  \begin{center}
    \begin{picture}(7.0,6.5)(0,0)
      \epsffile{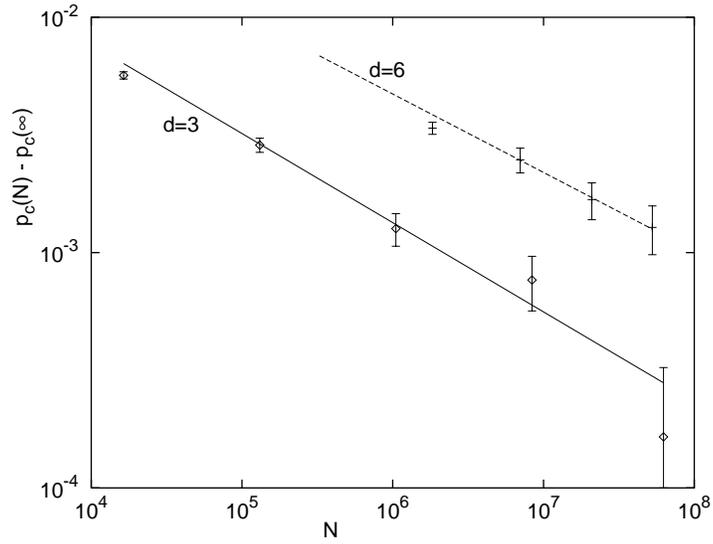}
    \end{picture}
  \end{center}
  \caption[]{The scaling of the percolation threshold with
             network size~$N=(d+1)L^d$.
             }
  \label{fig_scaling}
\end{figure}

\begin{figure}[htb]
  \begin{center}
    \begin{picture}(7.0,6.5)(0,0)
      \epsffile{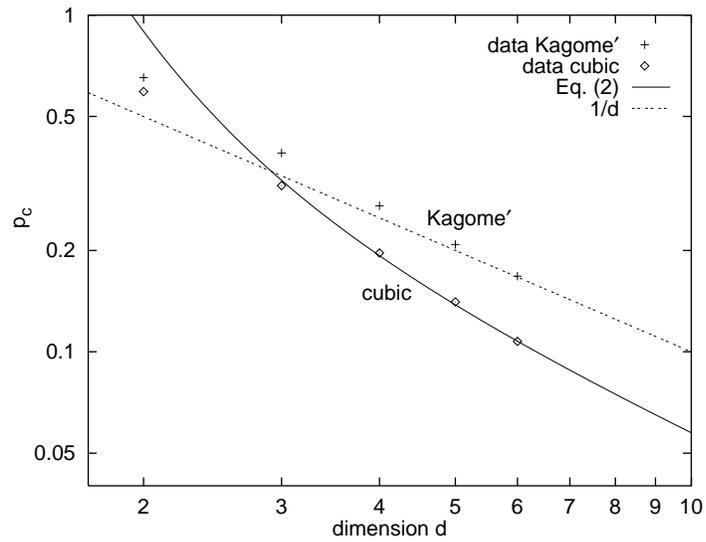}
    \end{picture}
  \end{center}
  \caption[]{The percolation thresholds as a function of
             dimension~$d$.
             Eq.~(\ref{eq_GSR})~is the result of
             Gaunt {\it et al.} (1976) for the thresholds of
             the cubic lattices.
             The thresholds for the Kagom\'e lattices
             scale as~$1/d$.}
  \label{fig_dimen}
\end{figure}

\begin{figure}[htb]
  \begin{center}
    \begin{picture}(10,2)(0,0)
      \epsffile{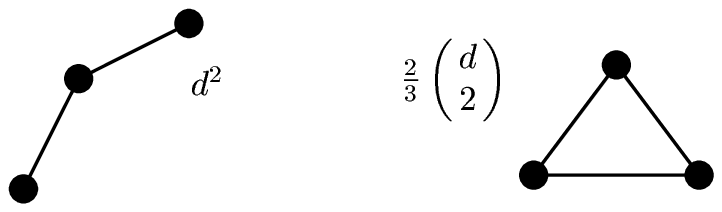}
    \end{picture}
  \end{center}
  \caption[]{The clusters of 3 sites.}
  \label{fig_count3}
\end{figure}

\begin{figure}[htb]
  \begin{center}
    \begin{picture}(10,2)(0,0)
      \epsffile{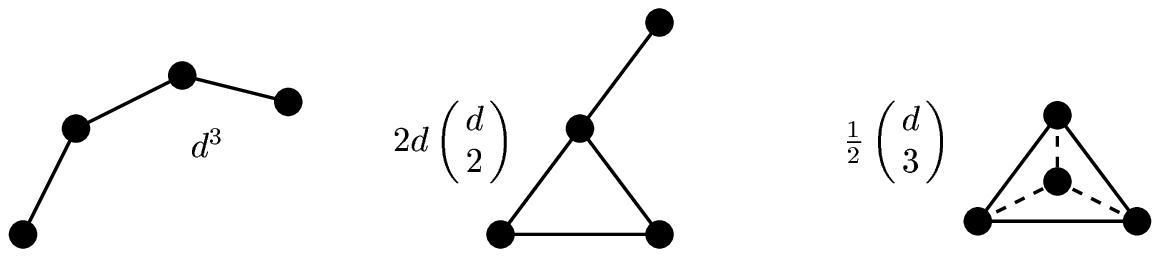}
    \end{picture}
  \end{center}
  \caption[]{The clusters of 4 sites.}
  \label{fig_count4}
\end{figure}

\begin{figure}[htb]
  \begin{center}
    \begin{picture}(10,4.3)(0,0)
      \epsffile{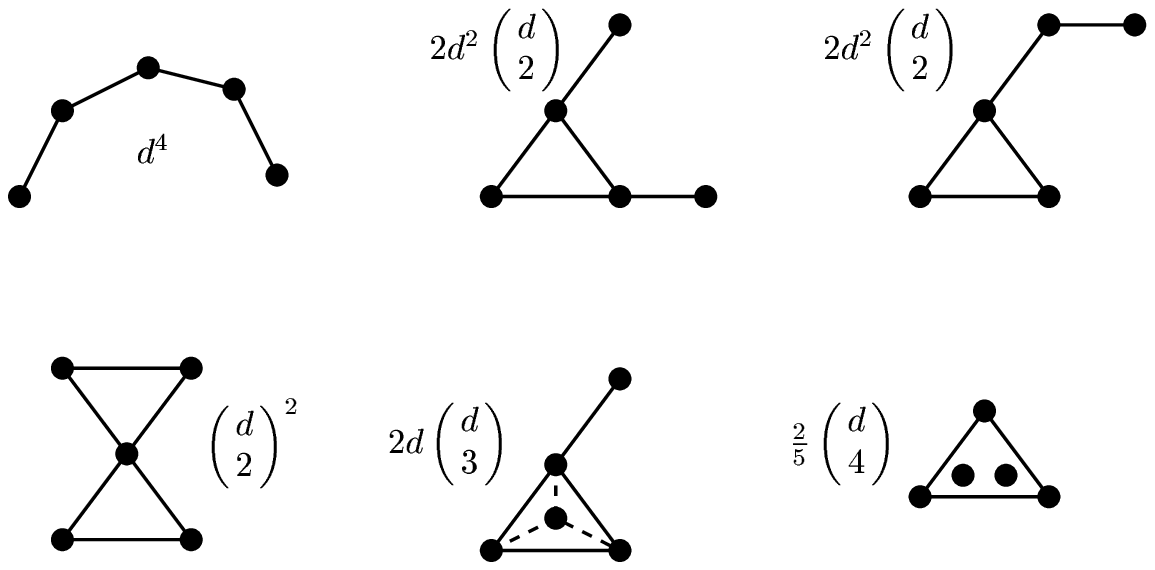}
    \end{picture}
  \end{center}
  \caption[]{The clusters of 5 sites.}
  \label{fig_count5}
\end{figure}

\begin{figure}[htb]
  \begin{center}
    \begin{picture}(10,4)(0,0)
      \epsffile{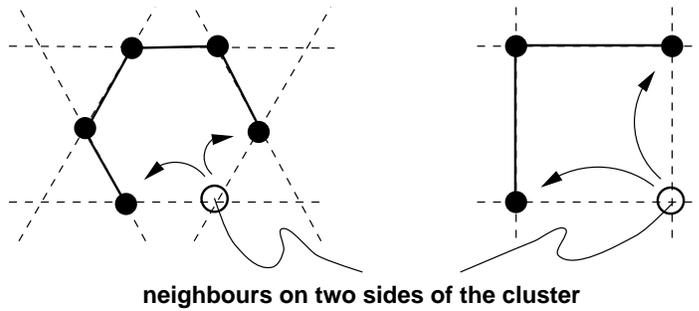}
    \end{picture}
  \end{center}
  \caption[]{On the Kagom\'e lattice the first instance of a
             cluster with fewer than usual neighbours occurs
             for clusters of five sites (left picture).
             On a cubic lattice it occurs for clusters
             of three sites (right picture).}
  \label{fig_snake}
\end{figure}

\begin{table}[htb]
  \begin{center}
      \caption[]{
           The site percolation thresholds of the
           cubic and Kagom\'e lattices in
           3, 4, 5 and 6 dimensions,
           as a function of the linear lattice size~$L$.
           The values for the cubic lattices given by
           Stauffer and Aharony (1992) are
           $0.3116$, $0.197$, $0.141$, and $0.107$.
           In the last row, marked `{\sc rw}', the values for
           the return probability of a random walker are given.
           The estimated error margins concerning the last digits
           are indicated between brackets.}
           \label{tab_big}
{\footnotesize
    \begin{tabular}{rccrccrccrcc}
    \hline
         \multicolumn{3}{c}{$d=3$} &
         \multicolumn{3}{c}{$d=4$} &
         \multicolumn{3}{c}{$d=5$} &
         \multicolumn{3}{c}{$d=6$} \\
         $L$ & cubic & Kagom\'e & $L$ & cubic & Kagom\'e &
         $L$ & cubic & Kagom\'e & $L$ & cubic & Kagom\'e \\
    \hline
      &          &          &  8& 0.2088(2)& 0.2787(2)&
     8& 0.1425(2)& 0.2080(3)&  6& 0.1043(2)& 0.1630(2)   \\

    16& 0.3233(2)& 0.3952(2)& 12& 0.2037(2)& 0.2753(2)&
    12& 0.1413(2)& 0.2076(2)&  8& 0.1049(2)& 0.1643(3)   \\

    32& 0.3171(2)& 0.3924(2)& 16& 0.2014(2)& 0.2736(2)&
    16& 0.1412(2)& 0.2078(2)& 10& 0.1060(2)& 0.1652(3)   \\

    64& 0.3139(2)& 0.3908(2)& 25& 0.1989(2)& 0.2726(2)&
    20& 0.1407(2)& 0.2080(2)& 12& 0.1064(2)& 0.1660(3)   \\

   128& 0.3125(2)& 0.3903(2)& 32& 0.1987(2)& 0.2724(2)&
    24& 0.1406(2)& 0.2081(2)& 14& 0.1067(2)& 0.1664(3)   \\

   250& 0.3119(2)& 0.3897(2)& 50& 0.1974(2)& 0.2719(2)&
    32& 0.1406(2)&          & 16& 0.1070(2)&             \\[2mm]

     $\infty$ & 0.3114(2) & 0.3895(2) &
     $\infty$ & 0.1967(3) & 0.2715(3) &
     $\infty$ & 0.1407(3) & 0.2084(4) &
     $\infty$ & 0.1079(5) & 0.1677(7)
   \\[2mm]
     {\sc rw} & 0.343(1)  & 0.417(1) &
     {\sc rw} & 0.195(1)  & 0.274(1) &
     {\sc rw} & 0.136(1)  & 0.208(1) &
     {\sc rw} & 0.105(1)  & 0.170(1) \\
    \hline
    \end{tabular}
}
  \end{center}
\end{table}

\end{document}